# Skyrmion based energy efficient straintronic physical reservoir computing

Md Mahadi Rajib[1], Walid Al Misba[1], Md. Fahim F. Chowdhury[1], Muhammad Sabbir Alam[1], Jayasimha Atulasimha[1,2*]

[1]Department of Mechanical and Nuclear Engineering, Virginia Commonwealth University, Richmond, VA 23384, USA

[2]Department of Electrical and Computer Engineering, Virginia Commonwealth University, Richmond, VA 23284, USA

*Email: jatulasimha@vcu.edu

Physical Reservoir Computing (PRC) is an unconventional computing paradigm, which exploits nonlinear dynamics of reservoir blocks to perform recognition and classification tasks. Here we show with simulations that patterned thin films hosting single/multiple skyrmions can implement energy efficient straintronic reservoir computing in the presence of room temperature thermal perturbation. This reservoir computing (RC) block is based on strain induced nonlinear breathing dynamics of skyrmions, which are coupled to each other through dipole and spin-wave interaction. The nonlinear and coupled magnetization dynamics is exploited to perform temporal pattern recognition. Two performance metrics, namely Short-Term Memory (STM) and Parity Check (PC) capacity are studied and shown to be promising (4.39 and 4.62 respectively), in addition to showing it can classify sine and square waves with 100% accuracy. This demonstrates the potential of such skyrmion based PRC. Furthermore, our study shows that nonlinear magnetization dynamics and interaction through spin-waves and dipole coupling have a strong influence on STM and PC capacity, thus explaining the role of physical interactions in a dynamical system on its ability to perform Reservoir Computing (RC).

## I. Introduction

Recently, reservoir computing (RC), originally developed from a Recurrent Neural Network (RNN) based framework [1, 2], has gained significant attention due to its faster and simpler processing of sequential or temporal data. In RC, the inputs are mapped into a high-dimensional space in the reservoir as shown in Fig. 1(a). Then, a one-dimensional readout is trained to extract the features of the high-dimensional pattern for classification. The primary advantage of RC is that, unlike an RNN, RC does not require training of the large complex reservoir; only training the output weights are sufficient for its successful operation. This straightforward and comparatively faster training approach has significantly reduced learning-cost compared to standard RNNs [3] which makes RC amenable to implement in power and hardware resource restrained edge devices that can learn in-situ [4].

Since the basic principle of a reservoir is to transform temporal/sequential inputs into a high-dimensional space and extract the features of the input by simple algorithms such as linear regression, any nonlinear dynamical system that exhibits a complex dynamic response to inputs can be utilized as a potential reservoir. Researchers have proposed various physical systems for implementing physical RC, including optical system [5, 6], memristors [7], and spintronic oscillators [8].

One such system is based on skyrmions. Skyrmions are topologically protected, localized, particle-like spin structures [9,10,11,12,13,14] as can be seen in Fig. 1(b). The requirement of very small depinning current compared to domain walls in racetrack memory devices [11, 15, 16, 17] and their scalability have made skyrmions a potential candidate for high density and low power spintronic devices [18, 19, 20]. For example, we have shown the feasibility of voltage control of skyrmions [21] and that skyrmion mediated voltage-controlled memory devices [22] can be scaled aggressively [23, 24]. Similarly, skyrmion can be a viable option for low-power, small-scale reservoir devices [1, 25]. Recently, D. Pinna et al. showed with simulations that random magnetic texture can be used as an ideal candidate for reservoir computing [26].

This was able to achieve simple temporal pattern recognition tasks by exploiting complex time varying resistance due to dynamics of skyrmions in response to an input AC voltage pulse.

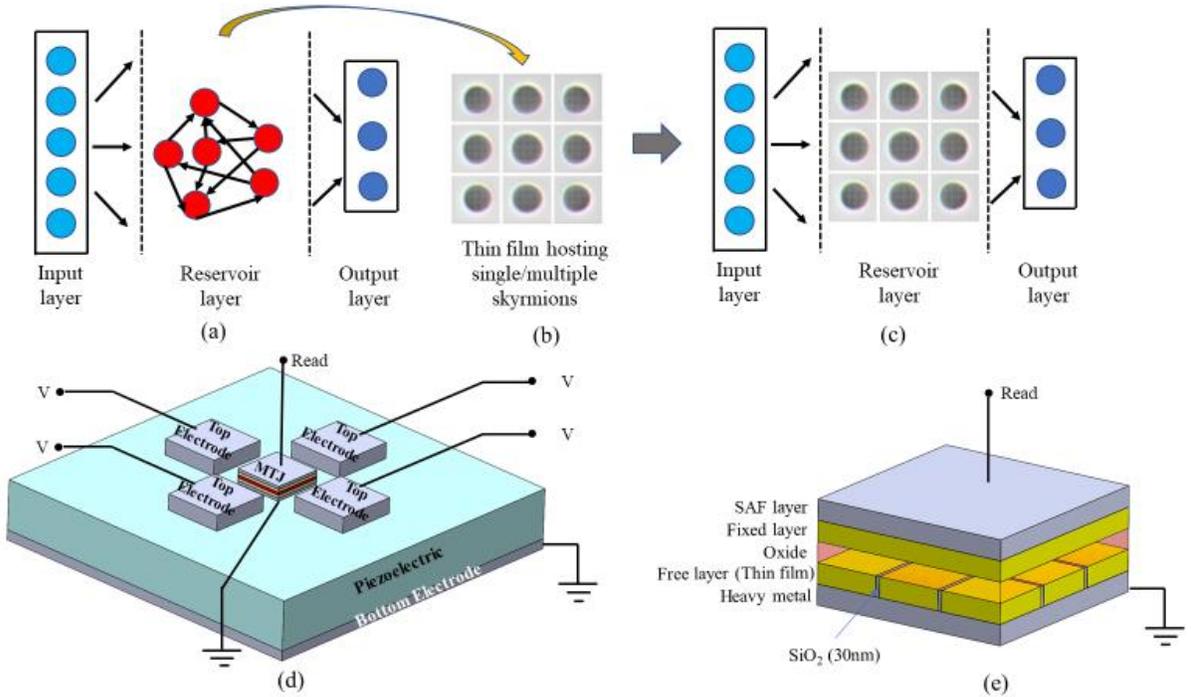

**Fig. 1.** (a) Schematic of a reservoir computing system based on a recurrent neural network, (b) Thin film hosting single/multiple skyrmions serves as a reservoir block, (c) A skyrmion based physical reservoir computing system which is obtained by replacing the recurrent network of (a) by the skyrmion-hosting thin film. (d) Proposed device where the free layer of the magnetic tunnel junction (MTJ) acts as the reservoir block. Voltage generated strain is provided as an input to the reservoir layer and the MTJ readout is used for reservoir computing output. (e) Different layers of MTJ stack.

However, many aspects of skyrmion based RC are unknown and yet to be explored for its successful implementation. In this paper, we utilize the nonlinear magnetization dynamics of skyrmions in response to the variation of perpendicular magnetic anisotropy (PMA) [27, 28] for physical RC. We applied voltage induced strain to manipulate the PMA of a thin film which in effect induces breathing of skyrmions and showed that an energy efficient straintronic skyrmion based physical reservoir computer can perform simple temporal pattern recognition task with 100% accuracy in the presence of room temperature thermal perturbation. Additionally, in the proposed configuration, the state of the skyrmion based reservoir computer block is amenable to reading out with magnetic tunnel junctions (MTJs) with enhanced sensitivity and reproducible indicating that the reservoir block can perform recognition task for multiple cycles. Furthermore, in this work, we use thin film hosting single/multiple skyrmions to study the *effect of coupling through dipole interaction and spin-waves on computing metrics* (STM and PC) and performing simple temporal pattern recognition tasks.

In section II, the reservoir model and method for providing an input to the reservoir block in the form of voltage generated strain as well as reading the output using MTJs is described. In section III, the Short-Term Memory (STM) and Parity Check (PC) capacity of the reservoir is investigated to assess the impact of spin-wave and dipole interaction on these metrics, thus studying the connection between the physics of coupling between these skyrmions and their computational performance. Section IV discusses the performance of reservoir computer in the presence of room temperature thermal noise, reproducibility of the output states from the reservoir, and method of enhancing sensitivity. Section V highlights the amount of energy dissipated for each input pulse while performing a simple pattern recognition task. Section VI concludes with a discussion on the future direction for this work.

## II. Reservoir model and simulation method

Our proposed device is shown in Fig. 1(d), which is made up of two main components: an MTJ stack and a piezoelectric substrate. The MTJ stack consists of a fixed layer, a tunnel barrier and a thin film hosting single/multiple skyrmions, which acts as a free layer. We consider heavy metal (HM)/CoFeB/MgO/CoFeB as the MTJ stack materials where the interface of HM/CoFeB and CoFeB/MgO gives rise to Dzyaloshinskii-Moriya interaction (DMI) and PMA respectively, which are two essential parameters for the formation and stabilization of magnetic skyrmions [29]. Four patterned electrodes on the piezoelectric substrate are positioned on four sides of the MTJ stack as shown in Fig. 1(d) to provide a voltage-generated localized strain large enough to control the thin film's magnetic anisotropy. When PMA is modulated by applying a voltage pulse between the electrodes on top and bottom of the piezoelectric material, skyrmion breathing is induced. The skyrmion core size varies due to the breathing of skyrmions which changes the magnetoresistance across the MTJ stack. This change in magnetoresistance is read out at regular interval by applying a read voltage pulse across the MTJ as shown in Fig. 1(d), which are then used for reservoir computing.

In this study, we considered ferromagnetic thin film of square geometry with 1000 nm, 1500 nm, and 2000 nm lateral dimensions for the implementation of the reservoir block. To demonstrate the effect of spin-wave interaction and dipole coupling on the performance of skyrmion based reservoir computer we formed two types of thin films: continuous and discontinuous, which is shown in Figs. 2(a)-(c). In continuous type of thin film, the breathing skyrmions can interact with each other through spin-wave and dipolar interaction whereas in discontinuous thin films, skyrmions can interact through dipolar interaction only since each of them is confined in a block of ~500nm×500nm dimension and separated by 30 nm gap (cells removed). Particularly, we placed single skyrmion in continuous/discontinuous thin film of all three lateral dimensions and four, nine, and sixteen skyrmions in thin film of 1000 nm, 1500 nm, and 2000 nm lateral dimension respectively. We note that continuous thin films are read as a single MTJ, while discontinuous thin films can also be read as a single MTJ with several blocks where the gap between the blocks in the free layer is filled with $SiO_2$ as shown in Fig. 1(e).

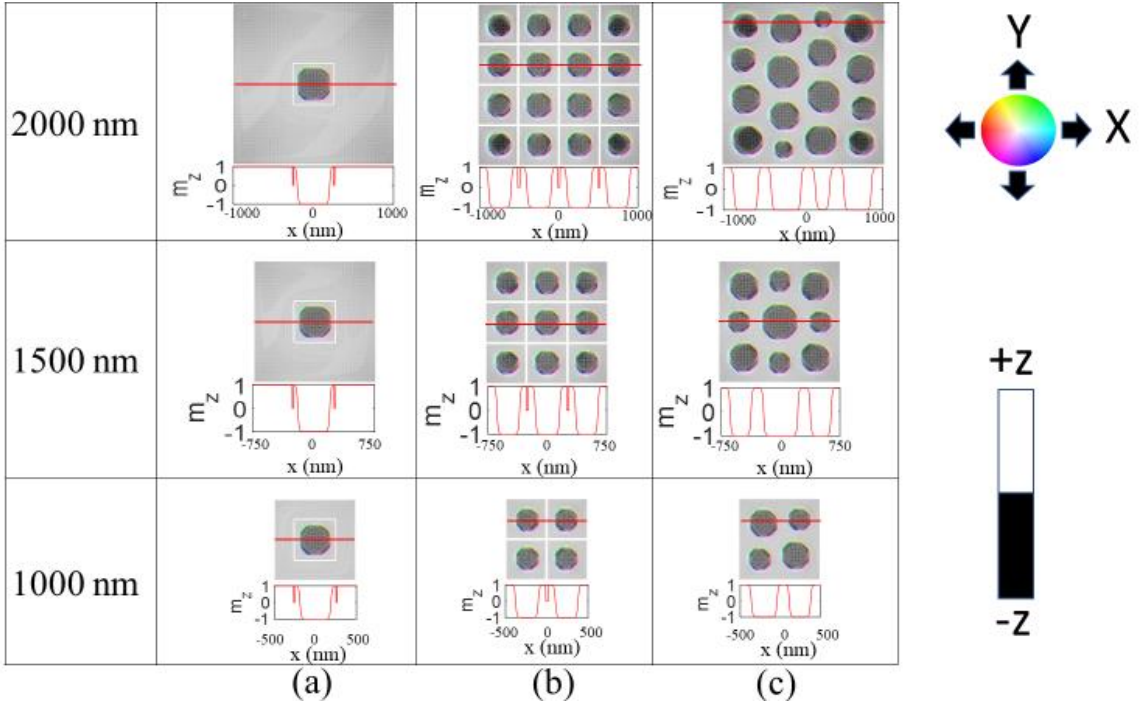

Fig. 2. Ferromagnetic thin film of 1000 nm, 1500 nm, and 2000 nm lateral dimension hosting (a) single skyrmion in a discontinuous thin film, (b) multiple skyrmions in a discontinuous thin film, (c) multiple skyrmions in a continuous thin film [The skyrmion states are confirmed by taking the profile (magnetization in the z-direction) along the radial direction (red line) of the skyrmion [30, 31]. We observe that for each of the skyrmions the magnetization profile looks like a $360^0$ Neel domain which verifies that the stabilized states are skyrmion states].

The magnetization dynamics of the breathing skyrmions induced by PMA modulation in the thin film is simulated by solving the Landau-Lifshitz-Gilbert (LLG) equation using the MᴜMᴀx3 simulation package [32]:

$$\frac{\partial \vec{m}}{\partial t} = \left( \frac{-\gamma}{1 + \alpha^2} \right) \left[ \vec{m} \times \vec{B}_{eff} + \alpha \{ \vec{m} \times (\vec{m} \times \vec{B}_{eff}) \} \right] \quad (1)$$

Where $\gamma$ and $\alpha$ represent the gyromagnetic ratio and the Gilbert damping coefficient respectively. $\vec{m}$ stands for the normalized magnetization vector, which is found by normalizing the magnetization vector ($\vec{M}$) with respect to saturation magnetization ($M_s$). The thin films are discretized into cells with dimensions of 2nm×2nm×1nm, which are well within the exchange length ($\sqrt{\frac{2A_{ex}}{\mu_0 M_s^2}}$). In equation (1), $\vec{B}_{eff}$ is the effective magnetic field, which includes the following components:

$$\vec{B}_{eff} = \vec{B}_{demag} + \vec{B}_{exchange} + \vec{B}_{DM} + \vec{B}_{anis} + \vec{B}_{stress} + \vec{B}_{thermal} \quad (2)$$

In equation (2), $\vec{B}_{demag}$ and $\vec{B}_{exchange}$ represent the effective field due to demagnetization energy and the Heisenberg exchange interaction respectively.

$\vec{B}_{DM}$ is the effective field due to Dzyaloshinskii-Moriya interaction which is defined as:

$$\vec{B}_{DM} = \frac{2D}{M_s} \left( \frac{\partial m_z}{\partial x}, \frac{\partial m_z}{\partial y}, -\frac{\partial m_x}{\partial x} - \frac{\partial m_y}{\partial x} \right) \quad (3)$$

Where D is the DMI constant and $m_x$, $m_y$, and $m_z$ are the components of unit magnetization vector $\vec{m}$ along x, y, and z direction, respectively.

$\vec{B}_{anis}$ is the effective field due to the perpendicular anisotropy and expressed by the following equation:

$$\vec{B}_{anis} = \frac{2K_{u1}}{M_s}(\vec{u}.\vec{m})\vec{u} \tag{4}$$

Where $K_{u1}$ is the first order uniaxial anisotropy constant with $\vec{u}$ a unit vector in the anisotropy direction.

By designing an electrode pair with a lateral dimension comparable to that of the thin film and a piezoelectric layer thickness equal to the lateral dimension of the thin film, a localized strain of sufficient magnitude to vary the magnetic anisotropy of the thin film can be created [33]. For an example, by applying 1 V between the bottom electrode and top electrodes of dimension 1200nm×1200nm each and taking 1000 nm thick piezoelectric Pb[Zr$_{0.52}$Ti$_{0.48}$]O$_3$ sufficient strain can be produced to modulate the required amount of PMA energy density in a 1000nm×1000nm thin film. In the micromagnetic simulation, the stress effect is modeled by the modulating $K_{u1}$.

Fluctuating thermal field is introduced with the following equation:

$$\vec{B}_{thermal} = \vec{\eta}(step)\sqrt{\frac{2\alpha k_B T}{M_s \gamma \Delta V \Delta t}} \tag{5}$$

where T is the temperature (K), $\Delta V$ is the cell volume, $k_B$ is the Boltzmann constant, $\Delta t$ is time step and $\vec{\eta}$ (step) is a random vector from a standard normal distribution which is independent (uncorrelated) for each of the three cartesian co-ordinates generated at every time step.

The following parameters for CoFeB are used for the simulation of magnetization dynamics of the skyrmions:

TABLE I. CoFeB material properties

| | |
|---|---|
| Saturation magnetization ($M_s$) | 1.3×10$^6$ A/m [34] |
| Exchange stiffness ($A_{ex}$) | 15×10$^{-12}$ J/m [34] |
| DMI | 0.65×10$^{-3}$ J/m$^2$ [29] |
| Thickness | 1 nm |
| Damping coefficient | 0.01 [35] |
| Perpendicular Magnetic Anisotropy constant | 1.0925×10$^6$ J/m$^3$ [36] |
| PMA modulation (△PMA) | 7.5×10$^3$ J/m$^3$ |

## III. Evaluation of short-term memory and parity check capacity

We analyzed the average magnetization in the z-direction of the thin films hosting single/multiple skyrmions generated in response to randomly distributed sine and square voltage pulses applied on the patterned electrodes and calculated the training and testing accuracy. Fig. 3 shows an example of PMA energy density variation with input wave forms to be classified and the corresponding average magnetization response of the reservoir in the z-direction from a discontinuous thin film of 1000 nm lateral dimension hosting a single skyrmion. The frequency of the input voltage pulse is considered to be 50 MHz.

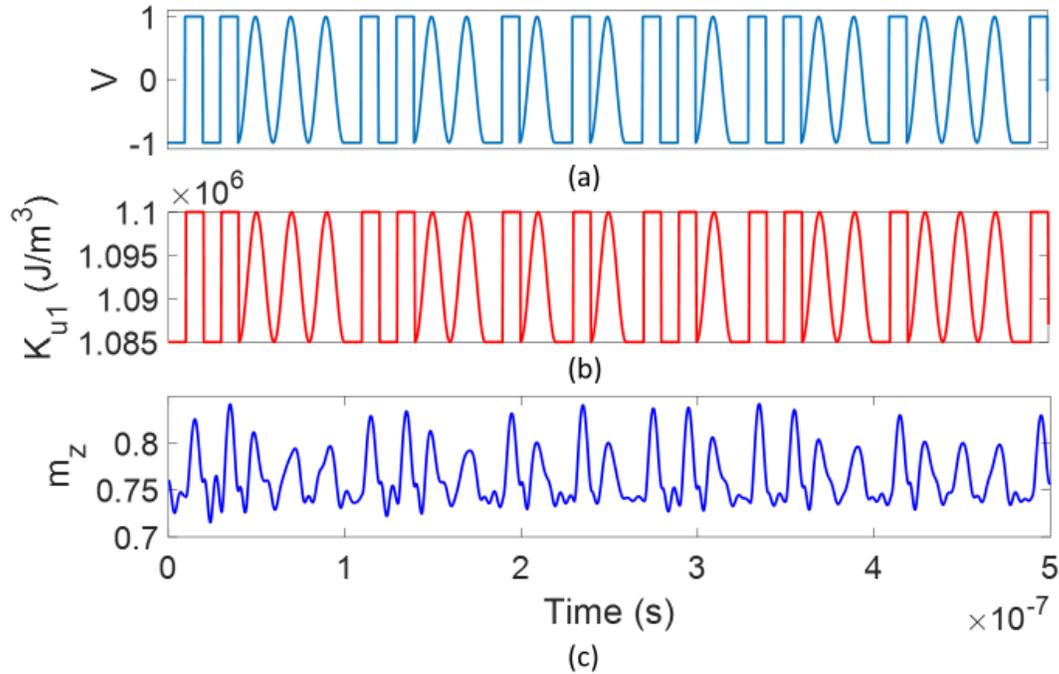

Fig. 3. (a) Input voltage pulse, (b) PMA modulation by voltage generated strain, (c) Change in average magnetization in the z-direction in response to PMA variation for a single skyrmion confined in a discontinuous thin film of 1000nm×1000nm dimension [From randomly distributed 200 sine and square wave pulses, response for the first 25 pulses is shown]

We observed that for all of the cases presented in this study, the training and testing accuracy are 100%. However, performance of different configurations as a reservoir block is estimated and compared on the basis of their STM and PC capacity. Fig. 4(a) and Fig. 4(b) respectively shows the STM and PC capacity for continuous/discontinuous thin film of 1000 nm, 1500 nm, and 2000 nm lateral dimension hosting single/multiple skyrmions for a maximum delay, $D' = 20$.

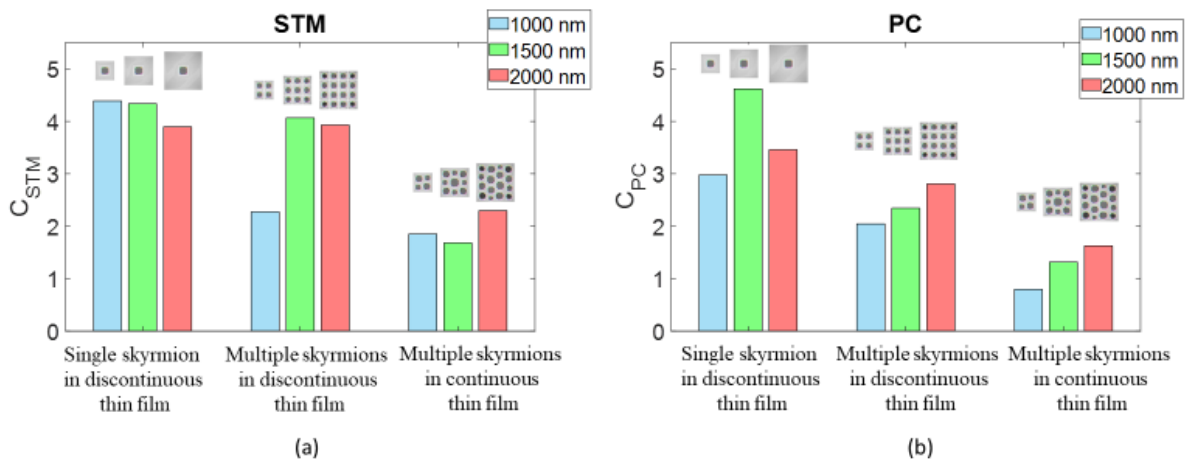

Fig. 4. (a) STM and (b) PC capacity of continuous/discontinuous thin films of 1000 nm, 1500 nm, and 2000 nm lateral dimension containing single/multiple skyrmions.

**Single skyrmion**

From Fig. 4 we can see that a single skyrmion located at the center of discontinuous thin film shows the highest STM and PC capacity ($C_{STM}$=4.39 and $C_{PC}$=4.62 for single skyrmion in 1000 nm and 1500 nm discontinuous thin film respectively). Before explaining the reason for a single skyrmion confined in a discontinuous film performing better than multiple skyrmions in continuous/discontinuous films, we first visualize the influence of the output of a reservoir block to an input on the STM and PC capacity with an example case of single skyrmion confined in a 1000nm×1000nm thin film. Memory capacity implies that the current magnetization response should be influenced by past pulse and due to the nonlinearity, the magnetization response should be different for different combination of past and present pulses. For this purpose, we observe the average magnetization response of thin film in the z-direction for four different past-present pulse combinations. Specifically, we investigate if the output of the present pulse is dependent on the past waveform by considering (a) Square-square, (b) Sine-sine, (c) Sine-square and (d) Square-sine combinations.

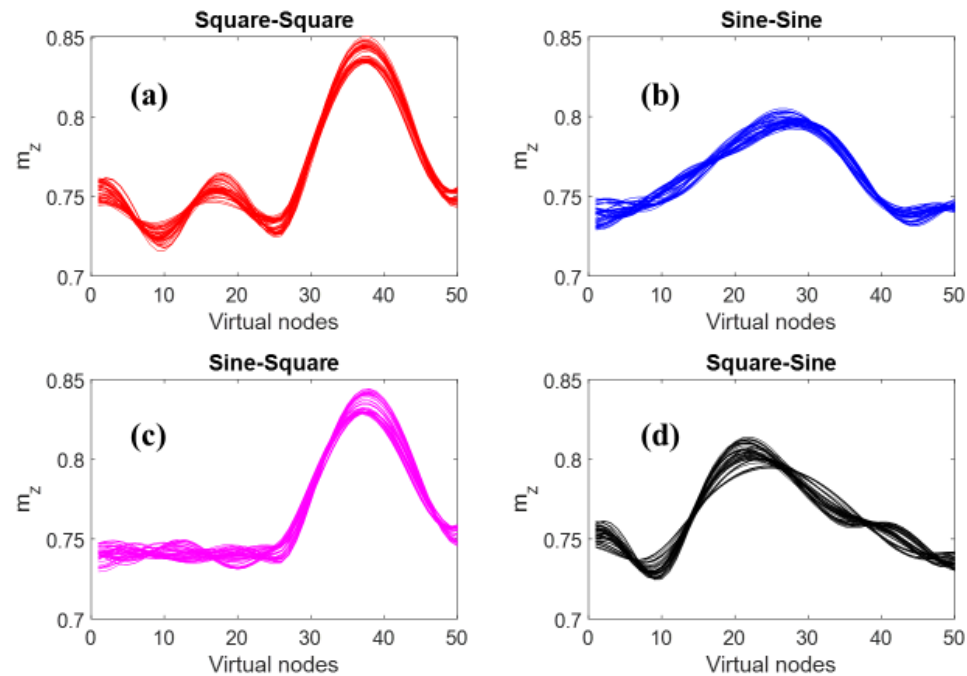

Fig. 5. Average magnetization response in the z-direction of the present pulse (second one in the combination) for the "past-present" pulse combination of (a) square-square, (b) sine-sine, (c) sine-square, and (d) square-sine for 50 virtual nodes.

Fig. 5 shows the average magnetization in the z-direction for present waveforms, which have a known past input and we can see that the magnetization responses are identical for same combination and different for different combination of waveforms. In Figs. 5 (a), 5(c) and Figs. 5(b), 5(d) the present pulses are square and sine respectively. When the square waveform is the present pulse, the magnetization has a big peak whereas for sine pulse the peaks are smaller. However, depending on the past input, the magnetization response takes different shapes while retaining the dominant feature arising from the present input. The difference in magnetization response for different combinations and similarity for the same combination suggests that the single skyrmion confined in a thin film has memory and nonlinearity.

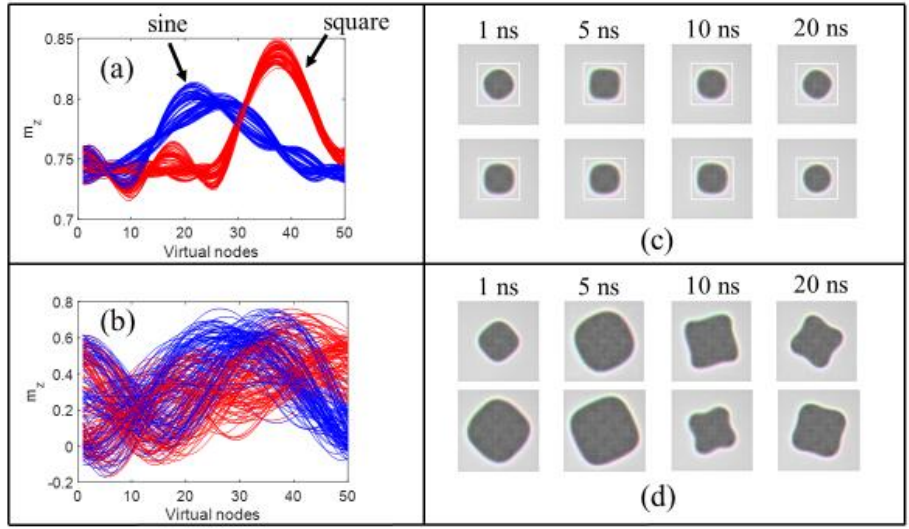

Fig. 6. Magnetization in the z-direction for a single skyrmion in a (a) discontinuous and (b) continuous thin film for different input pulses; red and blue color represent the magnetization response for square and sine input wave respectively. Magnetization states visited in response to two different square pulses for (c) discontinuous and (d) continuous thin film. (Top and bottom panel of (c) and (d) show the magnetization states while breathing in response to square pulses those are in 2nd and 12th position respectively in the randomly distributed pulse train).

We also estimated the STM and PC capacity for a single skyrmion hosted by a 1000nm×1000nm continuous thin film ($C_{STM}$=2.49, $C_{PC}$=1.27) and observed that these capacities are significantly smaller compared to single skyrmion in discontinuous thin film ($C_{STM}$=4.39, $C_{PC}$=2.98) of the same dimension. We note that though a single skyrmion can be stabilized in a 1000 nm×1000 nm continuous thin film but it cannot be stabilized in 1500nm×1500nm and 2000nm×2000nm thin film and therefore "single skyrmion in continuous thin film" case is not listed in Fig. 4. The difference in performance between the single skyrmion in a continuous (unconfined) and discontinuous (confined) thin film can be explained with Fig. 6. We see that for a single confined skyrmion, the magnetization response to sine and square wave inputs has visibly distinguishable features: sine and square wave inputs produce small and large peaks in average magnetization in the z-direction respectively (Fig. 6(a)). Particularly, the magnetization responses correspond to any of the four types of responses shown in Fig. 5. On the other hand, for a single skyrmion in a continuous thin film, these features are not visibly as identifiable as for the confined skyrmion (see Fig. 6(b) which is juxtaposed with Fig 6(a)). We take two instances of responses for square input and observe that the confined skyrmion is forced to breathe as a nearly circular skyrmion (Fig. 6(c)). Also, magnetization states visited while responding to the same kind of pulse are identical. On the contrary, the initially stabilized skyrmion in the unconfined geometry has neither circular nor identical shape for the same kind of input wave as shown in Fig. 6(d) and the shape of the skyrmion changes arbitrarily, which led to lower capacity of the reservoir to distinguish different input waves. Thus, we have correlated the magnetization dynamics (geometric shape of breathing) with the average magnetization along z-direction (affects the magnetoresistance read out) to the RC metrics (STM and PC) ultimately.

**Multiple Skyrmions**

From Fig. 4 we can see that multiple skyrmions in discontinuous thin film of 1000 nm, 1500 nm, and 2000 nm lateral dimension have higher STM and PC capacity than continuous thin film. Fig. 7 (a) shows the magnetization state visited at 1800 ns as an example for breathing dynamics of multiple skyrmions in continuous and discontinuous thin film. For continuous thin film, the skyrmions are slightly displaced and distorted whereas in discontinuous thin film skyrmion breathes retaining their nearly circular shape without being dislocated. The displacement and distortion in shape of skyrmions causes the continuous film reservoir to generate irregular responses (similar to single skyrmion in continuous thin film case) when excited with input waveforms of same sequence which in effect degrades the STM and PC capacity of continuous thin films hosting multiple skyrmions. Fig. 7 (b), (c) show an example case of magnetization response in the z-direction from a continuous and discontinuous thin film of 2000 nm lateral dimension hosting sixteen skyrmions respectively for the present pulse of "sine-square" wave sequences. We can see that the responses from continuous thin film vary more irregularly due to the distortion and displacement of skyrmions compared to discontinuous thin film which indicates that multiple skyrmion based discontinuous thin film performs better than continuous thin film.

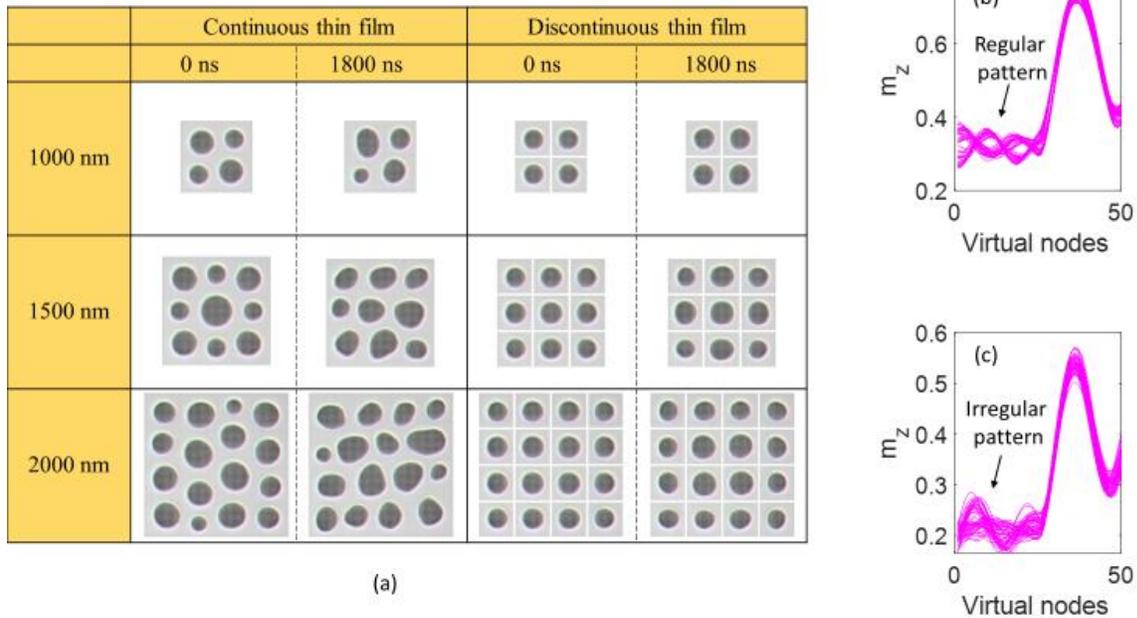

(a)

Fig. 7. (a) Magnetization states before applying input pulse (0ns) and after 1800 ns for symmetrically distributed multiple skyrmions in continuous and discontinuous thin film of 1000 nm, 1500 nm and 2000 nm lateral dimension. Average magnetization response in the z-direction of the present pulse (second one in the combination) from a (b) discontinuous and (c) continuous thin film of 2000 nm lateral dimension for the "sine-square" pulse combination.

## IV. A. Performance in the presence of room temperature thermal noise

From the non-thermal study, we observe that single skyrmion in a discontinuous thin film shows the highest STM and PC capacity and we further evaluate the performance of this RC in the presence of room temperature thermal perturbation. We increased the PMA energy from $1.0925 \times 10^6$ J/m$^3$ to $1.15 \times 10^6$ J/m$^3$ to stabilize the skyrmion with room temperature thermal noise as shown in Fig. 8(a). Also, to obtain responses from thermally stable skyrmion, the modulation of PMA energy was increased from $7.5 \times 10^3$ J/m$^3$ to $10 \times 10^3$ J/m$^3$. Other than these two values, all other parameters were kept as same as non-thermal case and we observed that single skyrmion in a discontinuous thin film can perform recognition task with 100% accuracy. However, the STM and PC capacity decreases ($C_{STM}$=1.6240, $C_{PC}$=0.7322) in the presence of thermal perturbation.

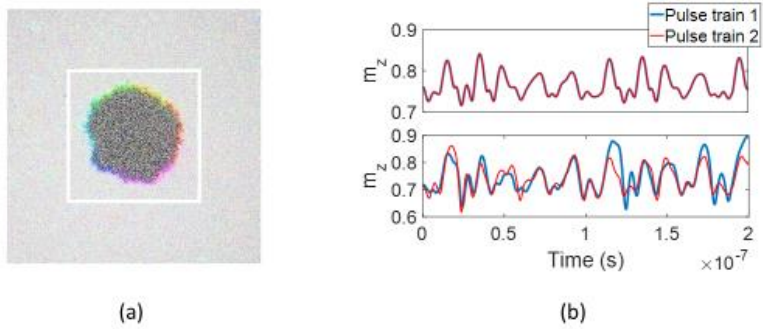

Fig. 8. (a) Single skyrmion in a discontinuous thin film stabilized at room temperature thermal perturbation, (b) change in magnetization in the z-direction for two pulse trains, each of those consisting of 200 randomly distributed sine and square wave pulses for non-thermal (top panel) and thermal (bottom panel) case [response for the first 10 pulses is shown].

## B. Reproducibility of the capacities

Before applying the input voltages, we first stabilize the skyrmion hosted in the thin film. We set single/multiple skyrmions in thin films of 1000 nm, 1500 nm, and 2000 nm lateral dimension and relax those for 500 ns to reach a stable state as shown in Fig. 2. Randomly distributed sine and square pulses are applied globally on these stable nanomagnets. Classification, memory capacity, and nonlinearity are evaluated based on the output (magnetoresistance) from the reservoir block. Once the randomly distributed waves are applied on the thin film, the skyrmion may dislocate or distort, which reduces the reproducibility of the output. Therefore, the skyrmion-hosting thin film is further stabilized for 500 ns after the application of pulse train and the reproducibility is checked. For single skyrmion in a 1000nm×1000nm thin film we can see in Fig. 8(b) that the magnetization in response to two sets of same voltage pulses are superposed (nearly superposed) for non-thermal (thermal) case, indicating the reproducibility of the capacities. Therefore, STM and PC capacities for two pulse trains are identical for non-thermal case but the capacities vary slightly in the presence of thermal noise. We note that the STM capacities are 1.624 and 1.4884 and PC capacities are 0.7322 and 0.4404 for first and second pulse train respectively in the presence of room temperature thermal noise. However, we believe that with an appropriate choice of material parameters, the STM and PC capacities, as well as reproducibility, can be improved in the presence of thermal perturbation.

## C. Enhancing the sensitivity

We have considered the average magnetization of the whole thin film for determining the readout and single skyrmion confined in a discontinuous thin film is shown to have the highest STM and PC capacity. In commercial applications where readability can be an issue, the output signal of the single skyrmion confined at the center of a discontinuous thin film can be enhanced by taking the average magnetization from the central region as shown in the Fig. 9 (magnetoresistance change is proportional to the change in average magnetization in the z-direction of the soft layer of a perpendicular MTJ).

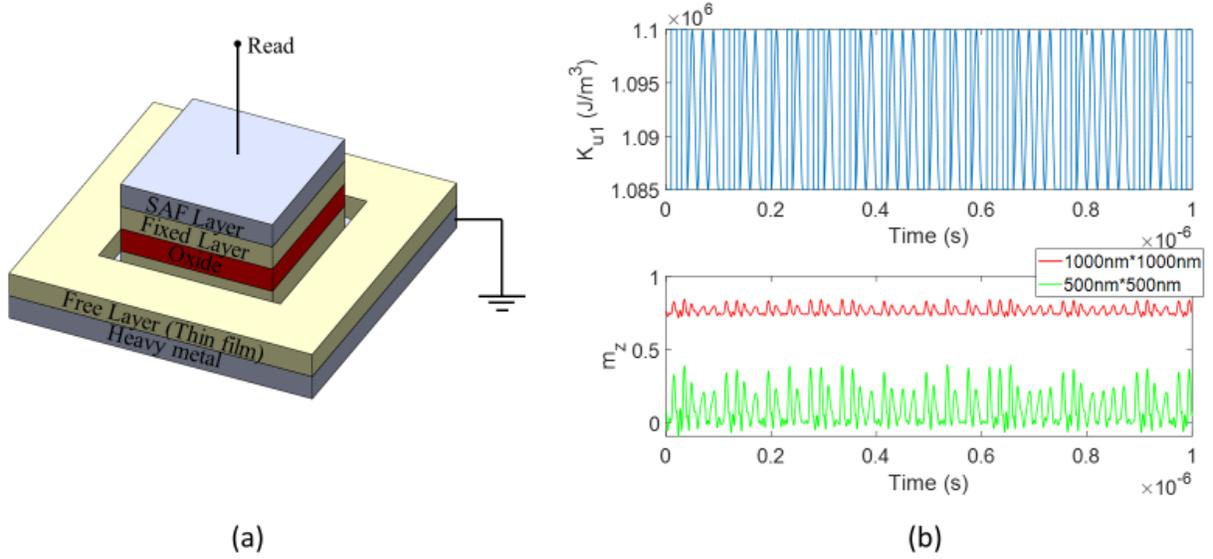

Fig. 9. (a) Read out of single skyrmion confined in a discontinuous thin film, (b) Enhancement of the magnetization response from the confined area where the single skyrmion is located.

## V. Energy dissipation

Energy is dissipated in two ways while using the proposed device to perform reservoir computing: reading the MTJ states and heating provided by charging current ($I$) running into the capacitor while generating strain to modulate PMA. The required maximum stress ($\sigma = \frac{\triangle PMA}{\frac{3}{2}\lambda_S}$) for the mentioned PMA modulation is 133 MPa considering CoFeB as the free layer material with saturation magnetostriction, $\lambda_S$=31 ppm [37] and Young's modulus, E=160 GPa [38]. This can be generated by a strain of $10^{-3}$, which can be developed by applying an electric field, E=1.0 MVm$^{-1}$ by designing four electrodes with a lateral dimension comparable to that of the thin film and a piezoelectric layer thickness equal to the lateral dimension of the thin film [33, 39]. Thus, the voltage applied to the top electrodes for a thin film of 1500 nm is, $V$= 1.5V. The piezoelectric layer between four top electrodes and a bottom electrode as shown in Fig. 1(d) create four capacitors in parallel where alternating voltage pulses are applied from a 1.5 V input. The frequency of the applied voltage pulses is 50 MHz and considering an input resistance of 50 Ω, the energy cost ($VItcos\theta$, where $\theta$ is the phase angle) for each input voltage pulse of 20 ns period ($t$) is 0.1fJ. For reading an MTJ state, less than 1fJ energy is required [40] and therefore for reading 50 states per input pulse, ~50fJ energy is dissipated. Thus, the total energy dissipated in a skyrmion based straintronic RC is ~50fJ per input pulse. We compare to the energy dissipated in skyrmion based RC with an equivalent CMOS-based echo-state network (ESN). The ESN can be simulated to replicate the performance of a skyrmion based RC [40]. The PC capacity of a single skyrmion hosted by a discontinuous thin film of 1500 nm lateral dimension is ~5, which is similar to the PC capacity of the spintronic nano-oscillator based RC and it has been shown that to achieve this PC capacity with CMOS-based ESN, ~300 pJ energy is required for each input pulse [40]. Thus, skyrmion based RC requires three-four orders of magnitude less energy compared to CMOS based ESN.

## VI. Discussion and Conclusion

Our work shows that the intrinsic nonlinear behavior of skyrmions in response to a voltage generated strain can be exploited for implementing extremely energy efficient reservoir computing. This skyrmion based straintronic physical reservoir computer can classify simple temporal inputs with 100 % accuracy in the presence of room temperature thermal perturbation.

Furthermore, we have studied the effect of the non-linear and coupled magnetization dynamics (breathing of skyrmions) on reservoir computing metrics (STM and PC) and specifically showed that for achieving high STM and PC, skyrmion(s) are required to be confined and retain nearly circular shape while breathing in response to input. Though it is expected that spin-wave and dipolar interaction between skyrmions would enhance the mapping ability of the reservoir block to high-dimensional space by adding complexity; we observe the opposite, the spin-wave and dipolar interaction deteriorate the performance of the reservoir computer. In multiple skyrmion-based continuous thin film there is only nonlinear interaction, which includes non-linear breathing dynamics of skyrmion, non-linearly varying dipolar interaction and non-linear spin-wave interaction. The presence of spin-wave interaction potentially displaces and distorts the skyrmions in continuous thin film and therefore the STM and PC capacity of multiple skyrmion based continuous thin film decreases. When the multiple skyrmions are separated by a gap in a discontinuous thin film, there is no spin-wave interaction and the nonlinearity in discontinuous film results from the nonlinear breathing of skyrmion and non-linearly varying dipolar interaction. Since there is no spin-wave interaction and due to the existence of boundary effect in discontinuous thin film, the skyrmions are not displaced and distorted and provide better STM and PC capacity compared to multiple skyrmions in continuous thin film. When a single skyrmion is placed at the center of a thin film and is separated by a gap from its ferromagnetic periphery, there remains the linearly varying dipolar interaction from the periphery and non-linear breathing of skyrmion. The coexistence of these linear and non-linear interactions makes a discontinuous thin film hosting a single skyrmion a "mixture reservoir" and this coexistence of linearity and non-linearity can be attributed to the highest performance provided by a single skyrmion in a discontinuous thin film in a similar way reported by M. Inubushi and K. Yoshimura [41].We note that single skyrmion performing better than multiple skyrmions is true for our case where input is applied simultaneously to all skyrmions. Applying inputs separately to different skyrmions may produce different results (while also being more challenging to fabricate) but this is beyond the scope of the current paper. We also observed that the skyrmion hosting thin film can go through multiple cycles since capacities are reproducible. Moreover, we showed that the output signal from a single skyrmion based reservoir can be enhanced. In summary, our work provides a pathway to realize highly energy efficient and high-performance physical reservoir computing using voltage induced strain driven breathing of skyrmions. Such a physical RC paradigm can be well suited to learning relatively simple tasks in-situ on edge devices where energy and hardware resources are severely constrained.

**Acknowledgement:** This work was funded by NSF SHF Small grant # 1909030 and Virginia Commonwealth Cyber Initiative (CCI) grants.

## Appendix A: Classification and STM/PC evaluation method

We apply a pulse train consisting of randomly distributed sine and square waves with 50 MHz frequency for 4µs to demonstrate simple temporal pattern recognition tasks. Therefore, there are 200 waves with 20-nanosecond periods each. From these waves, the first 160 waves are used for training and the remaining 40 waves are used for testing the reservoir computer. Randomly distributed sine and square pulse applied to the reservoir block which are labeled as 0 and 1 respectively for the classification task:

$$x(n) = \begin{cases} 0 \text{ for sine wave} \\ 1 \text{ for square wave} \end{cases} \tag{6}$$

$$x(n) = \{x_1, x_2, \ldots \ldots, x_T, x_{T+1}, \ldots, x_P\} = \{1,1,0, \ldots \ldots,0,1\} = \{Y_{Train}, Y_{Test}\} \tag{7}$$

where n is the index number of the input wave pulse and n ϵ {1,2,3, .......T, ...., P}. $Y_{Train}$ and $Y_{Test}$ represent the training and testing data:

$$Y_{Train} = \{x_1, x_2, \ldots \ldots \ldots, x_T\}$$

$$\text{And } Y_{Test} = \{x_{T+1}, x_{T+2}, \ldots \ldots \ldots, x_P\}$$

Where "P" denotes the total number of waves and "T" represents the number of waves for training.

The output (magnetization response) of the reservoir block can be represented by a N×P matrix:

$$r_{ij} = \begin{bmatrix} r_{11} & r_{12} & \ldots & r_{1P} \\ r_{21} & r_{22} & \ldots & r_{2P} \\ \ldots & \ldots & \ldots & \ldots \\ r_{N1} & r_{N2} & \ldots & r_{NP} \end{bmatrix} ; \tag{8}$$

i ϵ {1,2,3, ..........., N} and j ϵ {1,2,3, ..........., P}

N is the virtual node, which is N=$\frac{T}{\tau}$, where $T$ is time period of the waves and $\tau$ is the time interval at which the output (magnetoresistance of the MTJ stack) is measured.

The weight vector is obtained by:

$$W = Y_{Train} \times pinv(r_{ij}) \tag{9}$$

i ϵ {1,2,3, ..........., N} and j ϵ {1,2,3, ..........., T}as we consider the training data only.
Here *pinv* is the Moore-Penrose Pseudoinverse of matrix $r_{ij}$

The test data can be reconstructed by using the weight learnt during training:

$$Y_{RO} = W \times r_{ij} \tag{10}$$

i ϵ {1,2,3, ..........., N} and j ϵ {T+1, T+2, ..........., P} as we consider the testing data only.
The pattern recognition task determines whether these reconstructed outputs ($Y_{RO}$) can accurately predict the test data, $Y_{Test}$.

As previously stated, in addition to checking accuracy, the performance of a reservoir can be evaluated using STM and PC capacity. STM capacity characterizes the ability of a reservoir to reproduce the past input from the present output of the reservoir block. The input training and testing data with delay D ($Y_{Train,n-D}^{STM}$, $Y_{Test,n-D}^{STM}$) for short-term memory check is obtained as follows:

$$Y_{STM} = x(n-D) = \{Y_{Train,n-D}^{STM}, Y_{Test,n-D}^{STM}\} \tag{11}$$

On the other hand, PC capacity determines the nonlinear transformation capability of a reservoir computer. The input training ($Y_{Train,n-D}^{PC}$) and testing ($Y_{Test,n-D}^{PC}$) data for parity check is obtained with the help of modulo operation as follows:

$$Y_{PC} = [x(n\text{-}D) + x(n\text{-}D+1) + .... + x(n)] \bmod (2) = \{Y^{PC}_{Train,n-D}, Y^{PC}_{Test,n-D}\} \tag{12}$$

In the evaluation of STM and PC capacity, the ability of the reconstructed output to predict the test data with delay D is estimated from the following correlation coefficient [42]:

$$Cor(D) = \frac{\sum_{k=1}^{Z}(Y_{Test,n-D} - <Y_{Test,n-D}>)(Y_{RO,n-D} - <Y_{RO,n-D}>)}{\sqrt{\sum_{k=1}^{Z}(Y_{Test,n-D} - <Y_{Test,n-D}>)^2 \sum_{k=1}^{Z}(Y_{RO,n-D} - <Y_{RO,n-D}>)^2}} \tag{13}$$

Where $Y_{Test,n-D}$ and $Y_{RO,n-D}$ represent the test data and reconstructed output at delay D respectively and $<\cdots>$ denotes the mean value of "Z" number of data.

STM or PC capacity is estimated using the following equation [42]:

$$C_{STM/PC} = \sum_{D=1}^{D'} [Cor(D)]^2 \tag{14}$$

Where $D'$ is the maximum delay.

## Appendix B: STM and PC capacity comparison for variation in DMI and exchange stiffness

We observed that confined single skyrmions provide the highest STM and PC capacity regardless of thin film size. Here, we provide further STM and PC capacity of single/multiple skyrmion in a 1000 nm thin film of variable DMI and exchange stiffness. From Figs. 10 (a) and 10 (b), we can see that for DMI variation, a single skyrmion confined in a discontinuous thin film provides the highest STM and PC capacity. For the variation in exchange stiffness, the STM and PC are the highest for single skyrmions in a discontinuous thin film except at 14 pJ/m, where STM capacity for multiple skyrmions in a discontinuous thin film is slightly higher than single skyrmion in a discontinuous thin film as can be seen in Fig. 10 (c). However, the PC capacity for single skyrmion is much bigger than that of multiple skyrmions in discontinuous thin film (Fig. 10 (d)). These findings show that our observation of a single skyrmion having a greater performance capacity than multiple skyrmions is not limited to a set of optimum parameters.

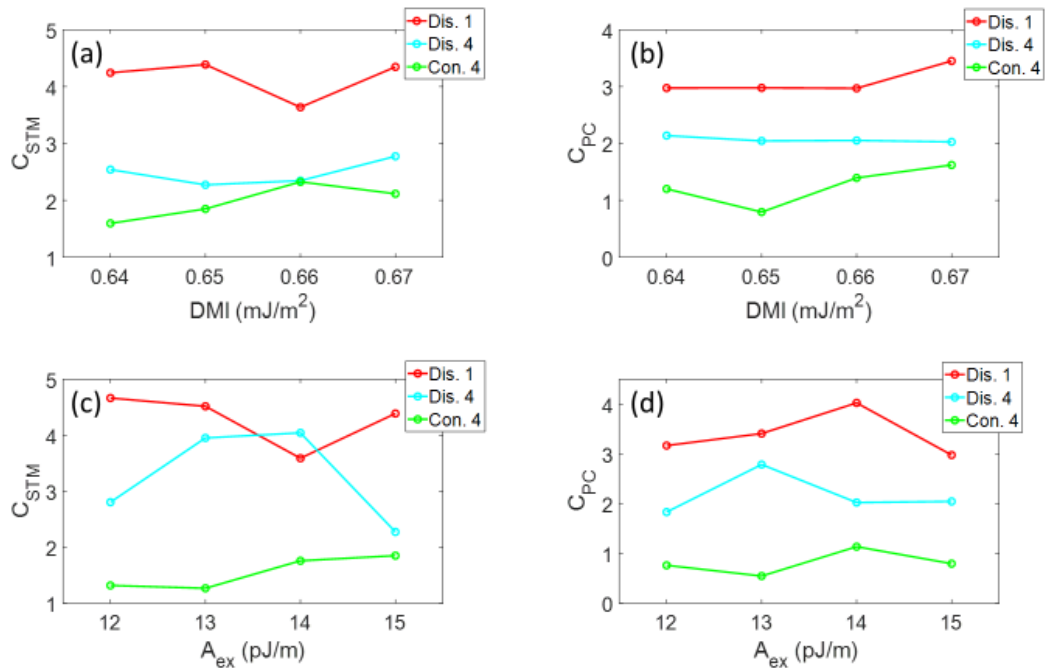

Fig. 10: STM and PC capacity for single/multiple skyrmions in a thin film of 1000 nm lateral dimension for variable (a, b) DMI and (c, d) exchange stiffness. (Dis. 1, Dis. 4 and Con. 4 represent single skyrmion in a discontinuous thin film, four skyrmions in a discontinuous thin film and four skyrmions in a continuous thin film respectively).

### References

[1] G. Tanaka, T. Yamane, J. B. Héroux, R. Nakane, N. Kanazawa, S. Takeda, H. Numata, D. Nakano, A. Hirose, Recent advances in physical reservoir computing: A review, Neural Networks 115 (2019).

[2] H. Jaeger and H. Haas, Harnessing nonlinearity: predicting chaotic systems and saving energy in wireless communication, Science 304, 78-80 (2004).

[3] Y. Zhong, J. Tang, X. Li *et al.* Dynamic memristor-based reservoir computing for high-efficiency temporal signal processing, *Nat Commun* **12,** 408 (2021).

[4] A. Morán, V. Canals, F. Galan-Prado *et al.* Hardware-Optimized Reservoir Computing System for Edge Intelligence Applications. *Cogn Comput* (2021).

[5] Y. Paquot, F. Duport, A. Smerieri *et al.* Optoelectronic Reservoir Computing. *Sci Rep* **2,** 287 (2012).

[6] K. Vandoorne, P. Mechet, T. Van Vaerenbergh *et al.* Experimental demonstration of reservoir computing on a silicon photonics chip. *Nat Commun* **5,** 3541 (2014).

[7] M. J. Marinella et al. Efficient reservoir computing with memristors. Nature Electronics 2, 437–438 (2019)

[8] J. Torrejon, M. Riou, F. Araujo *et al.* Neuromorphic computing with nanoscale spintronic oscillators. *Nature* **547,** 428–431 (2017).

[9] R. Takagi, Y. Yamasaki, T. Yokouchi *et al.* Particle-size dependent structural transformation of skyrmion lattice. *Nat Commun* **11,** 5685 (2020).

[10] U. Rößler, A. Bogdanov & C. Pfleiderer, Spontaneous skyrmion ground states in magnetic metals. *Nature* **442,** 797–801 (2006).

[11] A. Fert, V. Cros & J. Sampaio, Skyrmions on the track. *Nature Nanotech* **8,** 152–156 (2013).

[12] N. Romming et al. Writing and deleting single magnetic skyrmions. Science 341, 6146 (2013).

[13] S.-G. Je et al. Direct demonstration of topological stability of magnetic skyrmions via topology manipulation, ACS Nano 14, 3251-3258 (2020).

[14] A. Casiraghi, H. Corte-León, M. Vafaee *et al.* Individual skyrmion manipulation by local magnetic field gradients. *Commun Phys* **2,** 145 (2019).

[15] R. Tomasello et al. A strategy for the design of skyrmion racetrack memories. Sci. Rep. 4(1), 6784 (2014).

[16] J. Iwasaki, M. Mochizuki & N. Nagaosa. Universal current-velocity relation of skyrmion motion in chiral magnets. Nat. Commun. 4, 1463 (2013).

[17] J. Iwasaki, M. Mochizuki, & N. Nagaosa. Current-induced skyrmion dynamics in constricted geometries. Nat. Nanotechnol. 8, 742–747 (2013).

[18] X. Z. Yu et al. Skyrmion flow near room temperature in an ultralow current density. Nat. Commun. 3, 988 (2012).

[19] W. Jiang et al. Direct observation of the skyrmion Hall effect. Nat. Phys. 13, 162–169 (2016).

[20] S. Woo et al. Current-driven dynamics and inhibition of the skyrmion Hall effect of ferrimagnetic skyrmions in GdFeCo films. Nat. Commun. 9, 959 (2018).

[21] D. Bhattacharya et al. Creation and annihilation of non-volatile fixed magnetic skyrmions using voltage control of magnetic anisotropy. Nat. Electron. 3, 539–545 (2020).

[22] D. Bhattacharya & J. Atulasimha. Skyrmion-mediated voltage-controlled switching of ferromagnets for reliable and energy-efficient two-terminal memory. ACS Appl. Mater. Interface 10(20), 17455–17462 (2018).

[23] M. M. Rajib, W. A. Misba, D. Bhattacharya, F. Garcia-Sanchez, & J. Atulasimha. Dynamic skyrmion mediated switching of perpendicular MTJs: Feasibility analysis of scaling to 20nm with thermal noise. IEEE Trans. Electron Devices 67, 9 (2020).

[24] M. M. Rajib, W. A. Misba, D. Bhattacharya, & J. Atulasimha. Robust skyrmion mediated reversal of ferromagnetic nanodots of 20 nm lateral dimension with high Ms and observable DMI. Sci Rep 11, 20914 (2021).

[25] D. Prychynenko, M. Sitte, K. Litzius, B. Krüger, G. Bourianoff, M. Kläui, J. Sinova, & K. EverschorSitte, Magnetic Skyrmion as a Nonlinear Resistive Element: A Potential Building Block for Reservoir Computing, Phys. Rev. Applied 9, 014034 (2018).

[26] D. Pinna, G. Bourianoff, & K. Everschor-Sitte, Reservoir Computing with Random Skyrmion Textures, Phys. Rev. Applied 14, 054020 (2020).

[27] M. A. Azam, D. Bhattacharya, D. Querlioz and J. Atulasimha, Resonate and fire neuron with fixed magnetic skyrmions, Journal of Applied Physics 124, 152122 (2018).

[28] X. Chen, F. A. Araujo, M. Riou *et al.,* Forecasting the outcome of spintronic experiments with Neural Ordinary Differential Equations. *Nat Commun* **13,** 1016 (2022).

[29] R. Chen et al., Large Dzyaloshinskii-Moriya interaction and room-temperature nanoscale skyrmions in CoFeB/MgO heterostructures. Cell Reports Physical Science 2, 100618 (2021).

[30] H.-B. Braun, Fluctuations and instabilities of ferromagnetic domain-wall pairs in an external magnetic field. *Phys. Rev. B* **50**, 16485 (1994).

[31] X. S. Wang, H. Y. Yuan, & X. R. Wang. A theory on skyrmion size. *Commun Phys* **1,** 31 (2018).

[32] A. Vansteenkiste et al. The design and verification of MuMax3. AIP Adv. 4(10), 107133 (2014).

[33] J. Cui, J. L. Hockel, P. K. Nordeen, D. M. Pisani, C.-Y. Liang, and C. S. Lynch. A method to control magnetism in individual strain-mediated magnetoelectric islands. Appl. Phys. Lett. 103, 232905 (2013).

[34] A. Conca, J. Greser, T. Sebastian, S. Klingler, B. Obry, B. Leven, and B. Hillebrands. Low spin-wave damping in amorphous $Co_{40}Fe_{40}B_{20}$ thin films. Journal of Applied Physics 113, 213909 (2013).

[35] S. Iihama, S. Mizukami, H. Naganuma, M. Oogane, Y. Ando, and T. Miyazaki. Gilbert damping constants of Ta/CoFeB/MgO(Ta) thin films measured by optical detection. Phys. Rev. B 89, 174416 (2014).

[36] J. G. Alzate, P. K. Amiri, G. Yu, P. Upadhyaya, J. A. Katine, J. Langer, B. Ocker, I. N. Krivorotov, and K. L. Wang. Temperature dependence of the voltage-controlled perpendicular anisotropy in nanoscale MgO|CoFeB|Ta magnetic tunnel junctions. Appl. Phys. Lett. 104, 112410 (2014).

[37] Dexin Wang, Cathy Nordman, Zhenghong Qian, James M. Daughton, and John Myers. Magnetostriction effect of amorphous CoFeB thin films and application in spin-dependent tunnel junctions. Journal of Applied Physics 97, 10C906 (2005).

[38] Z. Tang et al. Magneto-mechanical coupling effect in amorphous $Co_{40}Fe_{40}B_{20}$ films grown on flexible substrates

[39] W. A. Misba, T. Kaisar, D. Bhattacharya and J. Atulasimha, Voltage-controlled energy-efficient domain wall synapses with stochastic distribution of quantized weights in the presence of thermal noise and edge roughness, IEEE Transactions on Electron Devices,1-9 (2021).

[40] M. F. F. Chowdhury, W. A. Misba, M. M. Rajib, A. J. Edwards, D. Bhattacharya, J. S. Friedman, and J. Atulasimha. Focused Surface Acoustic Wave induced nanooscillator based reservoir computing. arXiv:2112.04000 [cond-mat.mes-hall]

[41] M. Inubushi, K. Yoshimura. Reservoir Computing Beyond Memory-Nonlinearity Trade-off. Sci Rep 7, 10199 (2017)

[42] T. Yamaguchi, N. Akashi, K. Nakajima et al. Step-like dependence of memory function on pulse width in spintronics reservoir computing. Sci Rep 10, 19536 (2020).